\documentclass[letterpaper]{article} 
\usepackage{aaai2026}  
\usepackage{times}  
\usepackage{helvet}  
\usepackage{courier}  
\usepackage[hyphens]{url}  
\usepackage{graphicx} 
\usepackage{natbib}  
\usepackage{caption} 
\usepackage{algorithm}
\usepackage{algorithmic}

\usepackage{newfloat}
\usepackage{listings}
\DeclareCaptionStyle{ruled}{labelfont=normalfont,labelsep=colon,strut=off} 
\floatstyle{ruled}
\newfloat{listing}{tb}{lst}{}
\floatname{listing}{Listing}
\usepackage[subfigure]{tocloft}
\usepackage{subfigure}
\usepackage{bm,xcolor}
\usepackage{amsmath,amsfonts}
\usepackage{cleveref}

\DeclareRobustCommand\onedot{\futurelet\@let@token\@onedot}

\newcommand{\mc}[1]{\mathcal{#1}}
\newcommand{\mr}[1]{\mathrm{#1}}
\newcommand{\mb}[1]{\mathbf{#1}}

\newcounter{checksubsection}
\newcounter{checkitem}[checksubsection]

\title{MACS: Multi-source 
Audio-to-image Generation with Contextual Significance and Semantic Alignment}
\author{
    Hao Zhou\textsuperscript{\rm 1}\equalcontrib,
    Xiaobao Guo\textsuperscript{\rm 1}\thanks{Corresponding author.}\equalcontrib,
    Yuzhe Zhu\textsuperscript{\rm 1},
    Adams Wai-Kin Kong\textsuperscript{\rm 1}
}
\affiliations{
    \textsuperscript{\rm 1}
    College of Computing and Data Science, Nanyang Technological University, Singapore\\

    \{zhou0552,g240005\}@e.ntu.edu.sg, \{xiaobao.guo,adamskong\}@ntu.edu.sg
}

\usepackage{bibentry}

\begin{document}

\maketitle


\begin{abstract}
Propelled by the breakthrough in deep generative models, audio-to-image generation has emerged as a pivotal cross-modal task that converts complex auditory signals into rich visual representations. However, previous works only focus on single-source audio inputs for image generation, ignoring the multi-source characteristic in natural auditory scenes, thus limiting the performance in generating comprehensive visual content. To bridge this gap, we propose a method called MACS to conduct multi-source audio-to-image generation. To our best knowledge, this is the first work that explicitly separates multi-source audio to capture the rich audio components before image generation. MACS is a two-stage method. In the first stage, multi-source audio inputs are separated by a weakly supervised method, where the audio and text labels are semantically aligned by casting into a common space using the large pre-trained CLAP model. We introduce a ranking loss to consider the contextual significance of the separated audio signals. In the second stage, effective image generation is achieved by mapping the separated audio signals to the generation condition using only a trainable adapter and a MLP layer. We preprocess the LLP dataset as the first full multi-source audio-to-image generation benchmark. The experiments are conducted on multi-source, mixed-source, and single-source audio-to-image generation tasks. The proposed MACS outperforms the current state-of-the-art methods in 17 out of the 21 evaluation indexes on all tasks and delivers superior visual quality.
\end{abstract}

\begin{links}
    \link{Code}{https://github.com/alxzzhou/MACS}
\end{links}

\begin{figure}[ht]
    \centering
    \includegraphics[width=\linewidth]{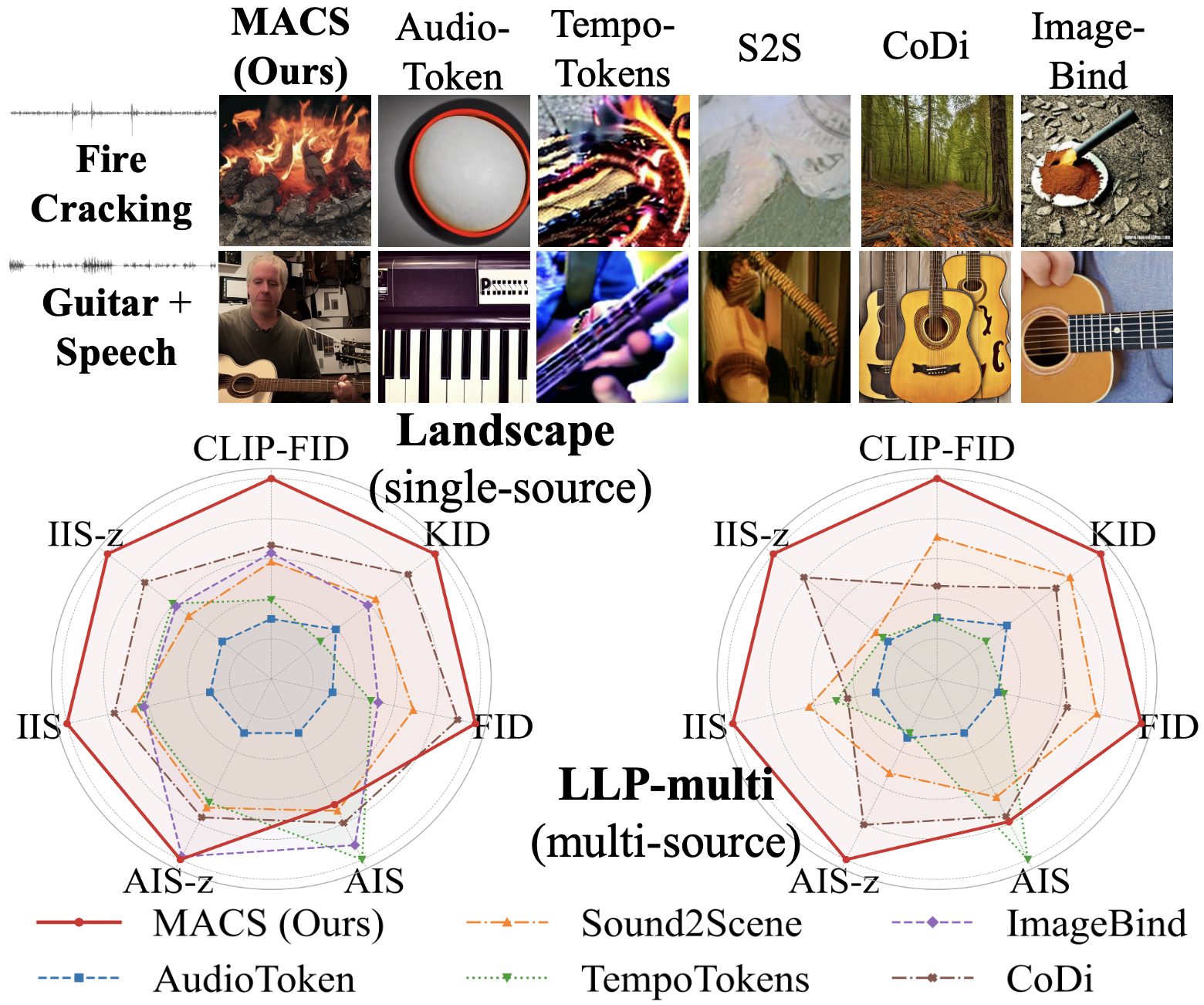}
    \caption{\textbf{Qualitative and Quantitative Comparison of MACS and Other SOTA Methods.} \textbf{Upper:} Generated images from single-source and multi-source audio datasets. \textbf{Lower:} Normalized radar maps (Left: single-source; Right: multi-source). FID, CLIP-FID, and KID are inverted.
    }
    \label{fig:teaser}
\end{figure}

\section{Introduction}
\label{sec:intro}

Audio-to-image generation has emerged as a cross-modal task that transforms rich and dynamic audio signals into semantically coherent visual representations. Early works in this area have demonstrated that audio cues, often rich with temporal dynamics and nuanced semantic information, can guide the synthesis of images~\cite{zhou2019talking,chen2017deep,duarte2019wav2pix,chatterjee2020sound2sight}. Recently, inspired by the success of diffusion models~\cite{ho2020denoising} and multimodal learning~\cite{radford2021learning,wu2023large,jia2021scaling}, more research works have demonstrated that models originally designed for text-to-image synthesis can be successfully adapted for audio inputs~\cite{yariv2023audiotoken,qin2023gluegen}, making it easier to develop an audio-to-image generation model. Audio-to-image generation is useful in many applications such as creative arts~\cite{lee2023generating}, multimedia content generation~\cite{girdhar2023imagebind}, and enhanced teaching and learning experiences by generating immersive visuals from sound by VR or AR systems~\cite{fitria2023augmented}.

Despite the recent success in conditioning image synthesis on audio, most of the existing literature focuses on single‐source audio inputs~\cite{yariv2023audiotoken,sung2023sound,qin2023gluegen}. 
In contrast, natural soundscapes are complex, often composed of multiple overlapping audio sources. Previous methods fall short on mixed audio signals; for example, they fail to effectively combine sounds like ``guitar and speech," resulting in incoherent images (see Fig.~\ref{fig:teaser}). These methods struggle to disentangle and utilize the full information embedded in real-world auditory scenes, limiting their ability to generate contextually comprehensive images.

To bridge the gap between multi-source audio and image generation, we propose \textbf{MACS}, the first framework that explicitly tackles multi-source audio-to-image generation. Our core idea is simple yet effective: ``separation before generation.” Rather than directly mapping complex mixtures to images, MACS first disentangles mixed audio into individual sources and then synthesizes an image that captures their combined semantics.
MACS is a two-stage framework that addresses three main challenges:
\textbf{1)} Mixed audio separation. Overlapping audio sources must be disentangled using robust separation techniques that preserve each source’s unique characteristics; \textbf{ 2)} Contextual significance and semantic alignment. The semantic content and relative importance of each separated source must be maintained and appropriately balanced to ensure coherent visual synthesis; \textbf{3)} Multi-source conditioning in diffusion. The system needs to map multiple concurrent audio representations to a single visual output using a diffusion model, where the overall scene can be generated effectively.

Specifically, we propose a multi-source audio separation model based on UNet~\cite{ronneberger2015u}. For semantic alignment, we project individual audio signals and their corresponding labels into the CLAP~\cite{wu2023large} space using a contrastive loss. Leveraging the pre-trained model \textit{provides additional prior knowledge} and \textit{enriches the semantic representation of the audio}. We also introduce a ranking loss to \textit{capture the contextual significance of each audio signal}. By disentangling the mixed audio from the physical environment, our approach enables the model to learn how to \textit{combine} these signals \textit{more effectively} for image generation. Analysis in Fig.~\ref{fig:attention} shows that separated audio embeddings can produce more localized and semantically aligned attention maps. In the second stage, the individual audio signals are transformed and mapped to a visual output through the diffusion process, where we use the trainable decoupled cross-attention module~\cite{ye2023ip} and an MLP layer for effective mapping while keeping the rest of the model frozen. 

To sum up, our contributions are as follows:
\begin{itemize}
    \item We propose MACS, the first audio-to-image framework that explicitly separates multi-source audio inputs.
    \item We propose to preserve the contextual significance and semantics of the separated audio signals by introducing a ranking loss and a contrastive loss in the CLAP space.
    \item We propose a scheme based on the decoupled cross-attention module to effectively merge multiple audio signals into a single image in the diffusion process.
    \item MACS outperforms the compared SOTA models on multi-source, mixed-source, and single-source audio-to-image generation tasks with significant margins.
\end{itemize}

\section{Related Works}
\label{sec:related works}

\subsection{Sound Source Separation}

Sound source separation aims to decompose a mixed audio signal into its constituent sound sources. Recently, researchers have pursued different training schemes, including supervised, unsupervised, and weakly-supervised methods to separate sound sources. Supervised models rely on large datasets with isolated ground-truth sources~\cite{wang2018supervised,luo2023music}, but are often limited to domains like speech and music, where clean source data is available. Unsupervised methods, such as PIT~\cite{yu2017permutation} and MixIT~\cite{wisdom2020unsupervised}, use unlabeled mixtures to learn representations, but require post-selection (e.g., a trained classifier) to identify separated sources. MixPIT~\cite{karamatli2022mixcycle} handles mixture-of-mixtures inputs directly but is limited in the number of separable sources. To overcome the limitations above, the weakly-supervised approaches explore large-scale mixture datasets and rely on high‐level semantic information rather than exact source ground truth for guidance~\cite{pishdadian2020finding}. However, most prior research works study vision- or text-conditioned audio separation~\cite{dong2023clipsep,mahmud2024weakly}. In this work, we follow the paradigm of the weakly-supervised methods and focus on leveraging versatile semantic information on unconditional sound separation.

\subsection{Multimodal Contrastive Pretraining and Audio-conditioned Image Generation}

Contrastive multimodal pretraining has become a cornerstone for learning aligned representations across text, image, and audio modalities~\cite{wang2023large,radford2021learning,chen2020uniter,kim2021vilt,jia2021scaling,li2023blip}. Pioneering works like CLIP~\cite{radford2021learning} train dual encoders with a contrastive loss on large-scale image-text pairs, enabling powerful cross-modal understanding. This framework has since been extended to other modalities. AudioCLIP~\cite{guzhov2022audioclip} and Wav2CLIP~\cite{wu2022wav2clip} introduce audio into the CLIP architecture, while CLAP~\cite{wu2023large} learns a joint embedding from over 600K audio-caption pairs for audio-text tasks. These aligned spaces facilitate diverse downstream applications, such as classification~\cite{wu2022wav2clip,guzhov2022audioclip}  and sound-guided image manipulation~\cite{lee2022sound,lee2024robust}.

Building on these embeddings, beyond the popular text-to-image generation~\cite{li2019controllable,reed2016generative,rombach2022high}, recent works explore audio-conditioned image generation~\cite{chatterjee2020sound2sight,oh2019speech2face}. Early attempts using GANs~\cite{wan2019towards} faced limitations in diversity and fidelity. More recent methods leverage diffusion models~\cite{rombach2022high} due to their superior stability and quality. For instance, AudioToken~\cite{yariv2023audiotoken} maps audio to discrete tokens compatible with Stable Diffusion, while ImageBind~\cite{girdhar2023imagebind} learns a unified embedding space for multiple modalities including audio, enabling generation via unCLIP~\cite{ramesh2022hierarchical}. Adapter-based diffusion models~\cite{mou2024t2i,ye2023ip} further enhance conditional generation with modular and effective conditioning.

While prior works in audio-to-image generation focus on single-source audio, we tackle the more complex and realistic setting of mixed audio inputs. Unlike methods that directly map entire audio scenes to images, we first separate audio sources before generation, allowing finer alignment between sound events and visual content. Our approach leverages the strong alignment capabilities of CLAP and diffusion models to enable image generation from rich, multi-source acoustic scenes.

\section{Methods}
\label{sec:method}

\begin{figure*}[ht]
    \centering
    \includegraphics[width=0.8\linewidth]{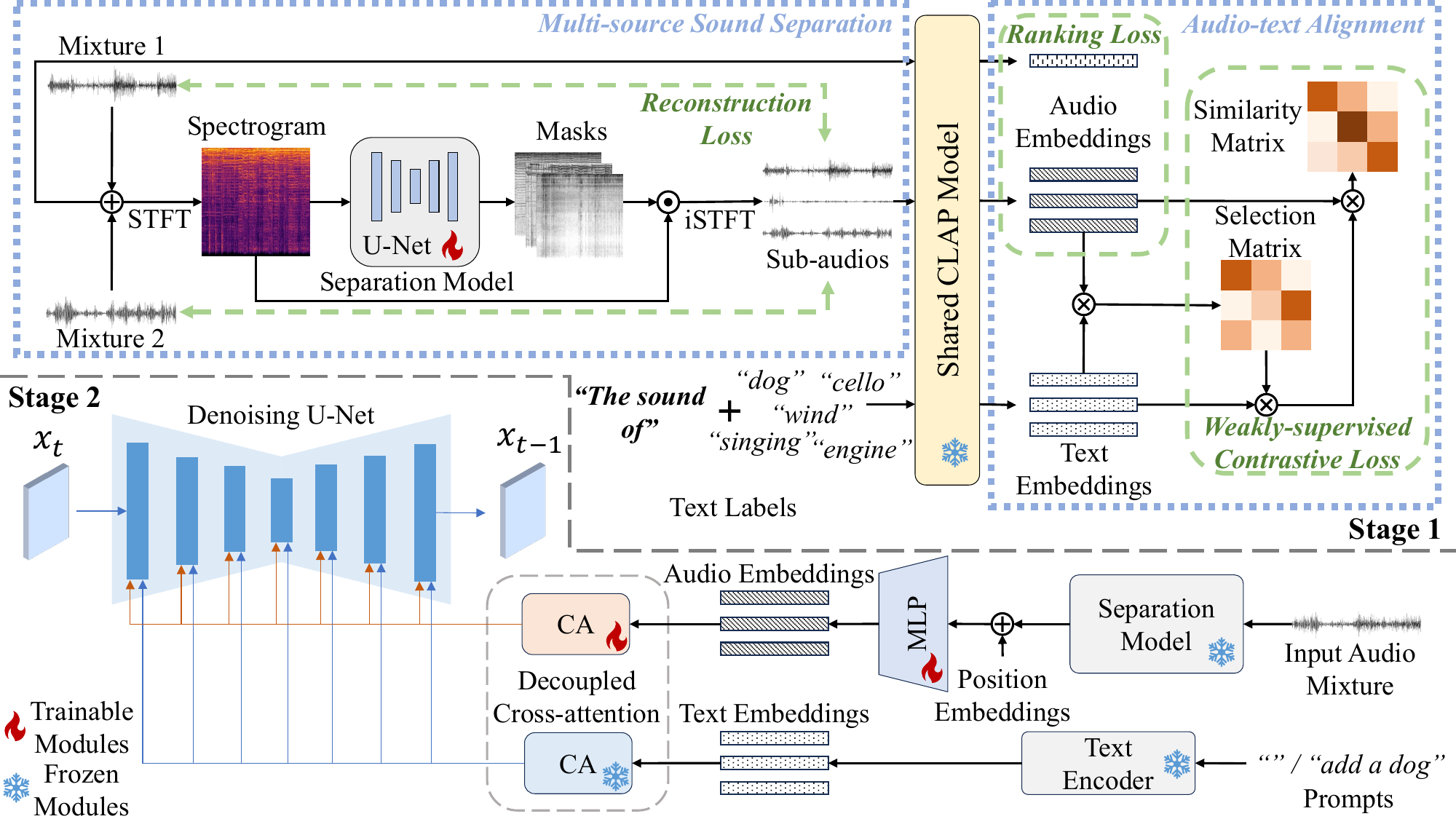}
    \caption{\textbf{An overview of the proposed two-stage MACS architecture.} \textit{Stage 1}: A Multi-source Sound Separation (MSS) model decomposes audio mixtures into sub-audios using reconstruction loss. The separated audios are embedded with CLAP, guided by contrastive and ranking losses to ensure audio-text semantic alignment and contextual significance. \textit{Stage 2}: A diffusion-based generator uses a decoupled cross-attention module to integrate audio embeddings and produce high-quality, semantically accurate images. MACS enables scalable MSS pre-training and image generation with fewer trainable layers.}
    \label{fig:model architecture}
\end{figure*}


The proposed MACS framework introduces a two-stage architecture for generating semantically meaningful images from multi-source audio.
By adopting a “\textit{separation before generation}” strategy, MACS effectively disentangles and recombines complex audio signals, leading to more accurate and expressive image generation. The brief description of MACS can be found in the caption of Fig.~\ref{fig:model architecture}. 

\subsection{Multi-source Audio Separation}
\label{sec:multi-source sound separation}

\paragraph{Problem Formulation}
Given a multi‑source audio mixture \(\mathbf{m}\), the separation model \(\mathcal{G}_{\theta}\) estimates \(M\) binary masks using a UNet \(\mathcal{U}_{\theta}\). First, \(\mathbf{m}\) is mapped to its spectrogram \(\mathcal{T}(\mathbf{m})\) via Short Time Fourier Transform (STFT), and the UNet predicts masks from the magnitude \(\lvert\mathcal{T}(\mathbf{m})\rvert\). Each mask is applied element‑wise to the magnitude, and the masked spectrograms are converted back to waveforms using the inverse STFT (iSTFT) with the original phase \(\phi(\mathcal{T}(\mathbf{m}))\). Formally,
\[
  \mathcal{G}_{\theta}(\mathbf{m})
  = \mathcal{T}^{-1}\!\Bigl(\lvert\mathcal{T}(\mathbf{m})\rvert \odot \mathcal{U}_{\theta}(\lvert\mathcal{T}(\mathbf{m})\rvert), \;\phi(\mathcal{T}(\mathbf{m}))\Bigr).
\]
The UNet operates on magnitude only, as phase is often unnecessary for many audio tasks~\cite{mahmud2024weakly}, and is used solely for waveform reconstruction.

\paragraph{Mixed Audio Separation}
Conventional audio separation trains on synthetic mixtures with single‑source ground truth, but often fails on real‑world audio. Inspired by MixIT~\cite{wisdom2020unsupervised}, we use an unsupervised Mixture of Mixtures (MoM) input and optimize a reconstruction loss to recover constituent sources. Like~\cite{dong2023clipsep,mahmud2024weakly}, we apply UNet‑based separation on spectrograms; however, our method is fully \emph{unconditional}, requiring no auxiliary inputs to the UNet.

Formally, given two mixtures $\bm{m}_1$ and $\bm{m}_2$, a new mixture of mixtures $\bm{m} = \bm{m}_1 + \bm{m}_2$ is formed. The separation model $\mc{G}_{\bm{\theta}}(\bm{m})$ produces $M$ separated signals \(\bm{S}=\{\bm{s}_1, \dots, \bm{s}_M\}\), whose sum reconstructs $\bm{m}$. To train the model, a reconstruction loss compares $\bm{S}$ and the original mixtures $\bm{m}_1$ and $\bm{m}_2$. Since \(M>2\) and source order is ambiguous, all possible bipartitions of \(\bm{S}\) are evaluated. The loss selects the partition $(\Lambda_1,\Lambda_2)$ that minimizes the total reconstruction error:
\begin{equation}
\begin{aligned}
      \mc{L}_{Rec}=\min_{(\Lambda_1,\Lambda_2)\in\Lambda}[&{\mc{L}_{SISDR}(\bm{m}_1, \sum_{i\in\Lambda_1}\bm{s}_i)}\\+
    &{\mc{L}_{SISDR}(\bm{m}_2, \sum_{i\in\Lambda_2}\bm{s}_i)}],
\end{aligned}
\end{equation}
where \(\Lambda\) represents the set of all non-empty, disjoint bipartitions of the index set \(A = \{1,2,\dots, M\}\):
\begin{equation}
    \Lambda=\{(\Lambda_1,\Lambda_2) \mid \Lambda_1\cup\Lambda_2= A, ~ \Lambda_1\cap\Lambda_2=\emptyset, ~ \Lambda_1,\Lambda_2\ne\emptyset\}.
\end{equation}
The reconstruction loss $\mc{L}_{SISDR}$ is measured by the negative scale-invariant signal-to-distortion ratio (SI-SDR)~\cite{le2019sdr}:
\begin{equation}
    \mc{L}_{SISDR}(\bm{m}_j, \hat{\bm{s}}_j)=-10\log_{10}\frac{\Vert\alpha \bm{m}_j\Vert^2_2}{\Vert\alpha \bm{m}_j-\hat{\bm{s}}_j\Vert_2^2},
\end{equation}
where $\hat{\bm{s}}_j = \sum_{i\in\Lambda_j}\bm{s}_i, j = \{1, 2\}$, and the scaling factor \(\alpha\) is $\frac{\hat{\bm{s}}_j^\top \bm{m}_j}{\Vert \bm{m}_j\Vert^2_2}$.
By minimizing the best-matching bipartition loss, the model learns to reconstruct the original mixtures without relying on fixed source assignments.

\subsection{Audio-text Alignment}
\label{sec:audio-text alignment}

While the separation model in \cref{sec:multi-source sound separation} captures rich audio features, its unsupervised training lacks the high‑level semantic alignment required for audio‑to‑image generation. To address this, we employ CLAP~\cite{wu2023large} to project separated signals, mixed audio, and text labels into a shared embedding space via an audio encoder \(\mathcal{P}_A\) and a language encoder \(\mathcal{P}_L\). We target two alignment objectives: (1) the contextual significance of each separated signal within the mixture, and (2) its semantic correspondence to its text label.

Specifically, let \(\mathcal{S}=[\mathbf{s}_1,\dots,\mathbf{s}_M]\) be the separated signals and \(\mathcal{T}=[t_1,\dots,t_{M'}]\) their labels. Each label \(t_i\) is prefixed with \textit{``The sound of''}; if \(M'<M\), we pad \(\mathcal{T}\) with \textit{``Noise''} to length \(M\). We then obtain
\[
  \mathcal{E}^A = \mathcal{P}_A(\mathcal{S}) \in \mathbb{R}^{M\times D}, 
  \quad
  \mathcal{E}^T = \mathcal{P}_L(\mathcal{T}) \in \mathbb{R}^{M\times D},
\]
where \(D\) denotes the embedding dimension. This procedure ensures both contextual and semantic alignment between audio signals and their textual descriptions.

\paragraph{Ranking Loss} 
We adopt a ranking loss to capture the contextual significance of separated audio sources within a mixture, helping the model identify which components are more semantically important. In real-world audio, some elements in a mixture carry greater importance. For example, background noise is often less relevant than distinct sounds like a dog bark or music. However, separation outputs are unordered, lacking inherent prioritization and making the ranking loss essential for guiding the model’s focus.

To enable our audio separation model to identify and prioritize more significant audio separations, we introduce a \textit{ranking loss}, formulated as:  
\begin{equation}
\label{eq:ranking loss}
    \mc{L}_{Rank} = 1 - r_s(\mb{S}, \mr{Sorted}(\mb{S})),
\end{equation}  
where \( r_s(\cdot,\cdot) \) represents the Spearman's rank correlation coefficient~\cite{spearman1987proof} quantifying the degree of discrepancy in the ranking of data between two arrays, and \( \mb{S} \in \mathbb{R}^M \) consists of the cosine similarities between the CLAP embedding of the original audio mixture, \( \mc{P}_A(\bm{m})\in\mathbb{R}^D \), and the separated audio embeddings, \( \mc{E}^A \in\mathbb{R}^{M\times D}\). Function $\mr{Sorted}(\cdot)$ outputs the sorted array in descending order. 
There is no strict requirement for the choice of sorting function. To ensure differentiability during training, we adopt the ranking optimization method from~\cite{blondel2020fast}, which allows direct optimization of ranking functions in deep models. This ranking loss guides the model to identify and prioritize important audio sources, enhancing its ability to preserve key semantic information.

\paragraph{Contrastive Loss} 
Our second focus is aligning separated audio signals with their text labels, which is crucial for image generation. To compensate for missing semantic cues in unsupervised learning, we apply a contrastive loss to align each audio with its corresponding text label.

Each separated audio \(\bm{s}_i\) is expected to semantically align with one of its associated text labels. However, explicit label information is not available at this stage, as the model receives only audio data as input. To address this, we perform a \textit{soft assignment} in a joint embedding space by aligning audio and text embeddings as follows:
\begin{equation}
    \mc{E'}^T = \mr{Softmax} \left( \frac{\langle \mc{E}^A \rangle \langle {\mc{E}^T} \rangle^\top}{\tau} \right) \mc{E}^T,\ \ \ \mc{E'}^T\in\mathbb{R}^{M\times D},
\end{equation} 
where \(\langle \cdot \rangle\) denotes \(L_2\) normalization, and \(\tau\) is a fixed temperature parameter set to 1e-2. 
This soft assignment ensures that each separated audio embedding is aligned with its most relevant text embedding, enabling semantic consistency without enforcing rigid one-to-one assignments. 

To align audio and text embeddings, we employ the contrastive loss~\cite{radford2021learning}:
\begin{equation}
\begin{aligned}
    \mc{L}_{CL}=&-\frac{1}{2M}\sum_{i=1}^M\log\frac{\exp(W_{ii})}{\sum_{j=1}^M\exp(W_{ij})}\\
    &-\frac{1}{2M}\sum_{i=1}^M\log\frac{\exp(W_{ii})}{\sum_{j=1}^M\exp(W_{ji})},
\end{aligned}
\end{equation}
where \(\tau'\) is a learnable temperature and $W$ computes cosine similarities,
\begin{equation}
    W = \frac{\langle \mc{E}^A \rangle \langle {\mc{E'}^T} \rangle^\top}{\tau'}\in\mathbb{R}^{M\times M}.
\end{equation}
This loss pulls matched audio–text pairs together and pushes mismatches apart, enhancing semantic alignment. 

The overall training objective for the first stage is:
\begin{equation}
\label{eq: stage1 loss}
    \mc{L}_1 = \lambda\mc{L}_{Rec} + \mu\mc{L}_{CL} + \gamma\mc{L}_{Rank},
\end{equation}  
where $\lambda$, $\mu$ and $\gamma$ are weights of the losses. 
The model is pre-trained on large-scale audio–text datasets, learning audio embeddings that transfer well to image generation.

\subsection{Multi-source Audio-to-image Generation}
\label{sec:multi-source audio-conditioned image generation}

To leverage text-to-image models such as Stable Diffusion, we adopt a decoupled cross‑attention module~\cite{ye2023ip}, originally designed as a lightweight adapter for image–text fusion, and extend it to handle multiple audio inputs. Given \(M\) audio embeddings \(\mathcal{E}^A \in \mathbb{R}^{M\times D}\), we first add trainable positional embeddings \(\mathcal{E}^{Pos}\in\mathbb{R}^{M\times D}\) and project the sum into the conditioning dimensionality \(D'\) via a multi‑layer perceptron (with layer normalization):
\begin{equation}
  \mathcal{E}'^A = \mathrm{MLP}\bigl(\mathcal{E}^A + \mathcal{E}^{Pos}\bigr) \in \mathbb{R}^{M\times D'}.
\end{equation}
For UNet query features \(\mathbf{H}\), the audio cross-attention is
\begin{equation}
  \mathbf{H}_{A} 
  = \mathrm{Softmax}\!\bigl(\tfrac{(\mathbf{H}\mathbf{W}_q)(\mathcal{E}'^A\mathbf{W}_k)^\top}{\sqrt{D'}}\bigr)\,(\mathcal{E}'^A\mathbf{W}_v),
\end{equation}
where \(\mathbf{W}_q\) is shared across modalities, and \(\mathbf{W}_k,\mathbf{W}_v\) are 
newly initialized. CLIP encodes the text prompt as \(\mathcal{E}^P\) and 
\begin{equation}
  \mathbf{H}_{T}
  = \mathrm{Softmax}\!\bigl(\tfrac{(\mathbf{H}\mathbf{W}_q)(\mathcal{E}^P\mathbf{W}'_k)^\top}{\sqrt{D'}}\bigr)\,(\mathcal{E}^P\mathbf{W}'_v).
\end{equation}
 Text conditioning is optional.
The combined output is
\begin{equation}
  \mathbf{H}' = \mathbf{H}_{A} + \mathbf{H}_{T}.
\end{equation}
In the second stage of training, only \(\mathbf{W}_k\), \(\mathbf{W}_v\), \(\mathcal{E}^{Pos}\), and the MLP parameters are updated, while the base model remains frozen. We optimize using the Stable Diffusion loss:
\begin{equation}
  \mathcal{L}_2 = \mathbb{E}_{\mathbf{z},\epsilon,t}\bigl\|\epsilon - \epsilon_\theta(\mathbf{z}_t, t, c)\bigr\|_2^2,
\end{equation}
where \(c = \{\mathcal{E}'^A,\mathcal{E}^P\}\). In our work, we extend the decoupled cross-attention module to audio inputs, enabling mapping multiple audio inputs into a single generated image.

\section{Experiments}
\label{sec:experiments}

\begin{table*}[ht]
\centering
\begin{tabular}{c|ccccccc}
\hline
\multicolumn{1}{c|}{Method} & FID$\downarrow$    & CLIP-FID$\downarrow$ & KID$\downarrow$    & AIS$\uparrow$    & AIS-z$\uparrow$  & IIS$\uparrow$    & IIS-z$\uparrow$  \\ \hline
ImageBind$^\star$~\cite{girdhar2023imagebind}    & \color{lightgray}{76.81}  & \color{lightgray}{21.17}    & \color{lightgray}{0.0088} & \color{lightgray}{0.0885} & \color{lightgray}{1.4219} & \color{lightgray}{0.6127} & \color{lightgray}{2.0361} \\
AudioToken~\cite{yariv2023audiotoken}    & 143.62 & 52.21    & 0.0431 & 0.0591 & 0.6201 & 0.4914 & 0.6799 \\
Sound2Scene~\cite{sung2023sound}         & \underline{105.14} & \underline{33.79}    & \underline{0.0240} & 0.0711 & 0.8176 & \underline{0.5545} & 0.7877 \\
TempoTokens~\cite{yariv2024diverse}      & 141.37 & 52.45    & 0.0494 & \textbf{0.0828} & 0.5932 & 0.5288 & 0.7259 \\
CoDi~\cite{tang2023any}                  & 116.67 & 44.96    & 0.0283 & 0.0747 & \underline{1.1068} & 0.5179 & \underline{1.4429} \\ \hline
AudioToken~\cite{yariv2023audiotoken} (w/ MSS)   & 130.77 & 47.03 & 0.0396 & 0.0633 & 0.6621 & 0.5173 & 0.6940 \\ \hline
\textbf{MACS}                           & \textbf{87.09} & \textbf{20.47} & \textbf{0.0157} & \underline{0.0754} & \textbf{1.3038} & \textbf{0.6269} & \textbf{1.7231} \\ \hline
\end{tabular}
\caption{Performance comparison with the baselines on LLP-multi (multi-source). The best results are \textbf{bold}, and the second-best results are \underline{underlined}. The method with a star$^\star$ is excluded for comparison but reference only.}
\label{tab:baselines llp}
\end{table*}

\begin{table*}[ht]
\centering
\begin{tabular}{c|ccccccc}
\hline
\multicolumn{1}{c|}{Method} & FID$\downarrow$    & CLIP-FID$\downarrow$ & KID$\downarrow$    & AIS$\uparrow$    & AIS-z$\uparrow$  & IIS$\uparrow$    & IIS-z$\uparrow$  \\ \hline
ImageBind$^\star$~\cite{girdhar2023imagebind}    & \color{lightgray}{41.69}  & \color{lightgray}{14.76}    & \color{lightgray}{0.0083} & \color{lightgray}{0.0892} & \color{lightgray}{1.1498} & \color{lightgray}{0.5808} & \color{lightgray}{1.7928} \\
AudioToken~\cite{yariv2023audiotoken}    & 102.85 & 40.68    & 0.0397 & 0.0663 & 0.5664 & 0.5426 & 0.6829 \\
Sound2Scene~\cite{sung2023sound}         & \underline{63.94}  & \underline{26.61}    & 0.0207 & 0.0725 & 0.7310 & 0.5445 & 0.6837 \\
TempoTokens~\cite{yariv2024diverse}      & 108.73 & 45.37    & 0.0510 & \textbf{0.0879} & 0.3863 & 0.5335 & 0.5153 \\
CoDi~\cite{tang2023any}                  & 70.20  & 31.49    & \underline{0.0206} & \underline{0.0789} & \textbf{1.0128} & 0.4920 & \underline{1.0869} \\ \hline
AudioToken~\cite{yariv2023audiotoken} (w/ MSS)   & 96.93 & 37.67 & 0.0305 & 0.0702 & 0.6218 & \underline{0.5479} & 0.7924 \\ \hline
\textbf{MACS}                          & \textbf{62.40} & \textbf{19.65} & \textbf{0.0142} & 0.0724 & \underline{0.8736} & \textbf{0.5532} & \textbf{1.1328}\\ \hline
\end{tabular}
\caption{Performance comparison on AudioSet-Eval (mixed-source). The best results are \textbf{bold}, and the second-best results are \underline{underlined}. The method with a star$^\star$ is excluded for comparison but reference only.}
\label{tab:baselines audioset}
\end{table*}

\begin{table*}[ht]
\centering
\begin{tabular}{c|ccccccc}
\hline
\multicolumn{1}{c|}{Method} & FID$\downarrow$    & CLIP-FID$\downarrow$ & KID$\downarrow$    & AIS$\uparrow$    & AIS-z$\uparrow$  & IIS$\uparrow$    & IIS-z$\uparrow$  \\ \hline
AudioToken~\cite{yariv2023audiotoken}    & 236.63 & 54.42    & 0.0402 & 0.0708 & 0.3527 & 0.6030 & 0.2900 \\
Sound2Scene~\cite{sung2023sound}         & 186.12 & 43.25    & 0.0280 & 0.1042 & 0.6519 & 0.6762 & 0.6368 \\
TempoTokens~\cite{yariv2024diverse}      & 212.69 & 50.70    & 0.0450 & \textbf{0.1251} & 0.6307 & 0.6703 & 0.8057 \\
ImageBind~\cite{girdhar2023imagebind}    & 207.93 & 41.49    & 0.0304 & \underline{0.1189} & \underline{0.8483} & 0.6673 & 0.7681 \\
CoDi~\cite{tang2023any}                  & \underline{158.31} & \underline{39.97}    & \underline{0.0180}  & 0.1094 & 0.6912 & \underline{0.6961} & \underline{1.0942} \\ \hline
AudioToken~\cite{yariv2023audiotoken} (w/ MSS)   & 202.54    & 50.57 & 0.0289    & 0.0817    &0.4351 &0.6161 & 0.3725\\ \hline
\textbf{MACS}                           & \textbf{147.23} & \textbf{26.91} & \textbf{0.0098} & 0.1015 & \textbf{0.8602} & \textbf{0.7422} & \textbf{1.4805}\\ \hline
\end{tabular}
\caption{Performance comparison on Landscape (single-source). The best results are \textbf{bold}, and the second-best results are \underline{underlined}. The evaluation is conducted on the standard train-test split~\cite{ruan2023mm}.}
\label{tab:baselines landscape}
\end{table*}

\subsection{Datasets}

\textbf{LLP-multi.} To address the lack of multi-source benchmarks, we construct LLP-multi from the LLP dataset~\cite{tian2020unified}, a subset of AudioSet~\cite{gemmeke2017audio}. We select videos with multiple labels and extract 6,595 frames with high audio-visual coexistence (6,314 with 2 labels, 242 with 3, and 35 with 4). LLP-multi captures concurrent audio-visual events with corresponding annotations, making it well-suited for \textit{multi-source tasks}.

\textbf{AudioSet-Eval.} AudioSet~\cite{gemmeke2017audio} contains over 2 million videos across diverse sound categories such as human voices, animal noises, and music. We use its evaluation split, filtering out ~20 poor-quality clips, resulting in 15,712 samples, 20.7\% with a single label and 79.3\% with multiple labels. With 610 distinct classes, this diverse set is used for \textit{ mixed-source evaluation}.

\textbf{Landscape.} The Landscape dataset~\cite{lee2022sound}, widely adopted in audio-to-image generation tasks, features 1,000 natural scene videos, each with one audio event from 9 classes. Following prior work~\cite{ruan2023mm}, we use a 90/10 train-test split for \textit{single-source evaluation}.

\textbf{FSD50K.} FSD50K~\cite{fonseca2021fsd50k} includes over 51,000 manually labeled audio clips across 200 classes from the AudioSet ontology. \textit{FSD50K is used to pre-train the audio separation model in the first stage of MACS.}

\subsection{Training Setup}
We use AdamW~\cite{loshchilov2019decoupled} in both stages with 
$\beta_1=0.9$, $\beta_2=0.999$, and a weight decay of 1e-2. Training uses a batch size of 16 (with gradient accumulation) on one RTX 4090D GPU. For LLP-multi and AudioSet-Eval, we report average results from 5-fold cross-validation. For Landscape, we follow the 90/10 train-test split in~\cite{ruan2023mm}. See Appendix G for more details.

\subsection{Evaluation Metrics} 
For a comprehensive performance evaluation, we gathered \textbf{seven metrics} for quantitative evaluation.  \textbf{(a)} the overall quality of the generated images including \textit{Fréchet Inception Distance (FID)}~\cite{heusel2017gans}, \textit{CLIP-FID}, and \textit{Kernel Inception Distance (KID)}~\cite{binkowski2018demystifying}, \textbf{(b)} pairwise similarity between generated images and ground truth images including \textit{Image-Image Similarity (IIS)}~\cite{yariv2023audiotoken} and \textit{IIS-z}, where ``z" means using z-score, and \textbf{(c)} pairwise semantic similarity between audio and images including \textit{Audio-Image Similarity (AIS)}~\cite{yariv2023audiotoken} and \textit{AIS-z}, ``z" for z-score.
More details are in Appendix E.

\subsection{Quantitative Analysis}
To ensure a comprehensive evaluation, we conducted experiments using \textit{multi-source, mixed-source, and single-source }audio datasets. We compared MACS with five state-of-the-art methods: AudioToken~\cite{yariv2023audiotoken}, Sound2Scene~\cite{sung2023sound}, TempoTokens~\cite{yariv2024diverse}, ImageBind~\cite{girdhar2023imagebind}, and CoDi~\cite{tang2023any}. Note that ImageBind and CoDi are competitive \textit{foundation models}, and we also tested these two on the evaluation datasets. Besides, since LLP-multi and AudioSet-Eval are included in ImageBind’s pre-training dataset~\cite{girdhar2023imagebind}, its performances in Tab.~\ref{tab:baselines llp} and Tab.~\ref{tab:baselines audioset} are presented in gray for \textit{reference only}.  The results are averaged over 5-fold cross-validation.

\textbf{Multi-source Audio.} We benchmarked MACS against five SOTA methods on \textbf{LLP-multi} (see Tab.~\ref{tab:baselines llp}), a \textit{fully} multi-source audio dataset. MACS significantly improves image quality (left three metrics) and further enhances content fidelity and semantic consistency across multiple sources (right four metrics).
We contend that the ``\textit{separation before generation}” strategy is key to its performance compared with methods that condition directly on mixed audio. Moreover, MACS achieves competitive results and outperforms ImageBind on two metrics, underscoring its strong capability in multi-source audio-to-image generation.

\textbf{Mixed-source Audio.} 
We further evaluate MACS on the mixed-source AudioSet-Eval dataset, which is more challenging than LLP-multi with over 600 event classes.
Despite the increased complexity, MACS consistently generates high-quality images. As shown in Tab.\ref{tab:baselines llp} and Tab.\ref{tab:baselines audioset}, overall image quality improves on mixed-source datasets. Unlike baselines that overlook audio mixing, MACS handles both the single- and multi-source inputs, making it versatile and scalable to an arbitrary number of audio sources.

\textbf{Single-source Audio.} 
MACS also performs strongly on Landscape, achieving state-of-the-art results on most metrics (Tab.~\ref{tab:baselines landscape}) and outperforming the next-best method by a substantial margin.
Compared to leading baselines and two strong foundation models, the audio separation process in MACS enhances single audio quality by minimizing noise; thereby, the generated images can be both higher in quality and semantic relevance.

\textbf{Multi-source Sound Separation (MSS) is Adaptable.} To evaluate MSS, we integrated MACS stage 1 outputs into AudioToken~\cite{yariv2023audiotoken}, which transforms audio clips into embeddings concatenated with the prompt ``\textbf{A photo of a...}'' for Stable Diffusion. We extended it to produce \(M\) audio tokens from the \(M\) separated signals, denoting this variant as “AudioToken (w/MSS)” (see Tab.~\ref{tab:baselines llp}–\ref{tab:baselines landscape}). Compared to the original, MSS markedly enhances performance, showing MACS’s effectiveness and adaptability.

\textbf{Audio Separation Model is Flexible.} We evaluated model performance using an alternative separation mechanism by replacing our module with MixIT and assessing generation quality on the LLP‑multi dataset (Tab.~\ref{tab:comparison}). Although MixIT operates at the waveform level, our spectrogram‐based separator outperforms it on every metric, underscoring its advantages. Nevertheless, MixIT remains competitive, demonstrating the audio‐separation component’s versatility within the MACS framework.

\begin{table*}[ht]
\centering
\begin{tabular}{c|ccccccc}
\hline
Sep. Model         & FID$\downarrow$   & CLIP-FID$\downarrow$ & KID$\downarrow$    & AIS$\uparrow$    & AIS-z$\uparrow$  & IIS$\uparrow$    & IIS-z$\uparrow$  \\ \hline
MixIT~[49] (Waveform)   & 98.73 & 28.42    & 0.0201 & 0.0688 & 1.0471 & 0.5782 & 1.3819 \\ \hline
MACS (Spectrogram) & \textbf{87.09} & \textbf{20.47}    & \textbf{0.0157} & \textbf{0.0754} & \textbf{1.3038} & \textbf{0.6269} & \textbf{1.7231} \\ \hline
\end{tabular}
\caption{MACS benchmarked with MixIT on LLP-multi.}
\label{tab:comparison}
\end{table*}

\begin{figure}[ht]
    \centering

\includegraphics[width=\linewidth]{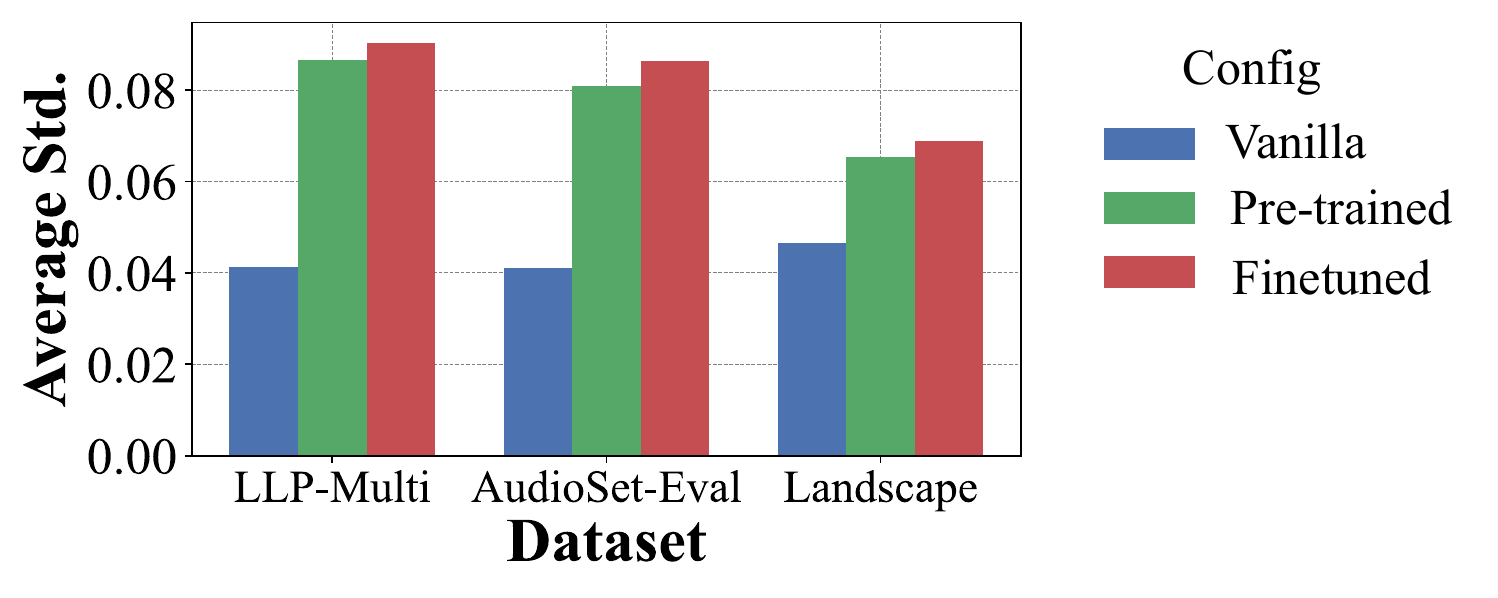}
    \caption{The average standard deviation of the similarity scores between the text labels of each audio mixture and its separations across three datasets.}
    \label{fig:std}
\end{figure}

\textbf{Pre-training Facilitates Multi-source Sound Separation.} We pre-trained the Multi‐source Sound Separation (MSS) model on FSD50K in stage 1 to enrich its audio representations. To assess the impact of pre‐training on separation performance, we evaluated three configurations (see Fig.~\ref{fig:std}): (1) \textbf{Vanilla}, using only reconstruction loss; (2) \textbf{Pre‐trained}, our default model with reconstruction and audio–text alignment losses (ranking and contrastive); and (3) \textbf{Fine‐tuned}, the pre‐trained model further trained for 10 epochs on the target dataset. We measured semantic alignment by computing the cosine similarity between $M'$ (mixture text-label embeddings) and $M$ (separated-audio embeddings), and reported the average standard deviation on the test set. Results demonstrate that adding semantic alignment loss during pre‐training substantially enhances separation quality, with fine-tuning providing comparable gains. These findings indicate that robust MSS pre‐training alone can suffice for audio‐to‐image generation, potentially eliminating the need for further fine-tuning.

\begin{figure}[ht]
    \centering
    \includegraphics[width=\linewidth]{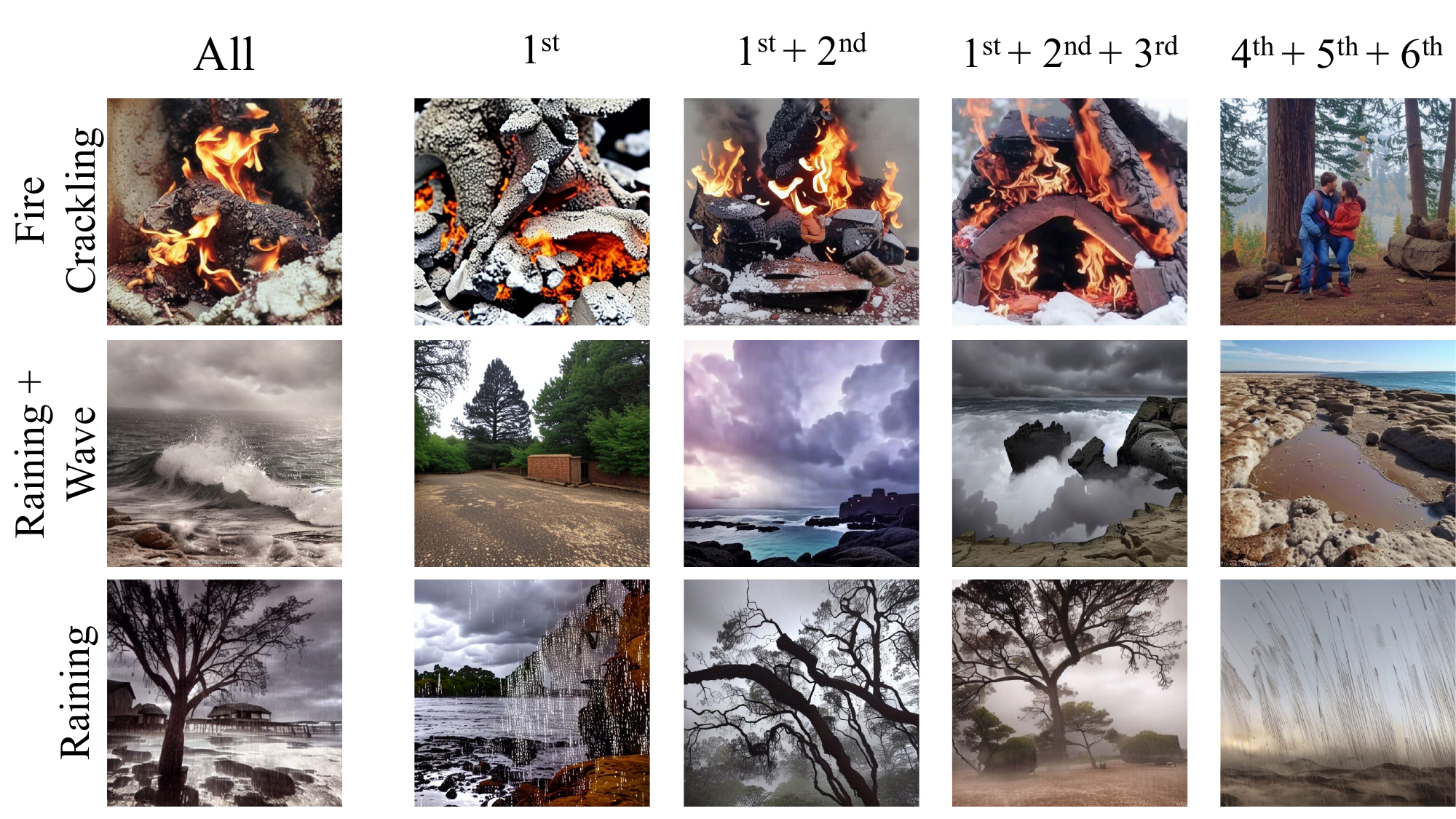}
    \caption{\textbf{Ranking loss helps sort the contextual significance.} Embeddings in higher ranking contain more important semantic information for accurate image generation.}
    \label{fig:tokens}
\end{figure}

\subsection{Qualitative Results}
\textbf{MACS produces higher quality visuals.} The upper part of Fig.~\ref{fig:teaser} presents images generated from single‐ and multi‐source audio using MACS and baseline methods. MACS consistently produces more realistic and semantically aligned images. For instance, generating vivid flames in the \textit{``Fire Crackling''} category, whereas baselines often produce abstract visuals. For multi‐source audio, MACS reliably captures the expected scenes, while methods such as CoDi and ImageBind frequently fail to reflect the audio content. See more qualitative examples in Appendix~C.

\textbf{Ranking Loss Helps Contextual Importance Learning.} The ranking loss in Eq.~\ref{eq:ranking loss} trains the model to capture the contextual importance of audio signals for image generation. With \(M=6\), we evaluated five configurations: all embeddings; only the first; the first two; the first three; and the last three. As shown in Fig.~\ref{fig:tokens}, higher‑ranked embeddings represent more salient audio events, with the first three encoding most of the semantic information. In single‑source cases, the first embedding alone suffices to generate semantically accurate images (see the first and last rows).

\begin{figure}[htbp]
	\centering
	\begin{subfigure}
		\centering
		\includegraphics[width=0.18\columnwidth]{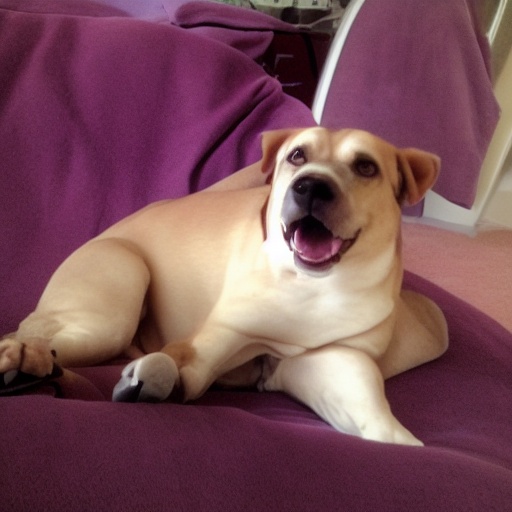}
	\end{subfigure}
	\centering
	\begin{subfigure}
		\centering
		\includegraphics[width=0.18\linewidth]{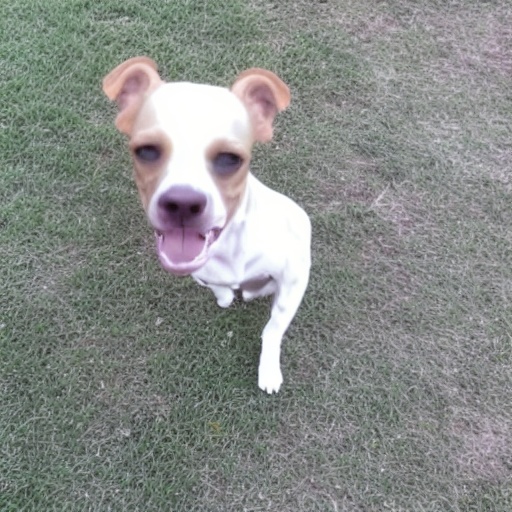}
	\end{subfigure}
	\centering
	\begin{subfigure}
		\centering
		\includegraphics[width=0.18\linewidth]{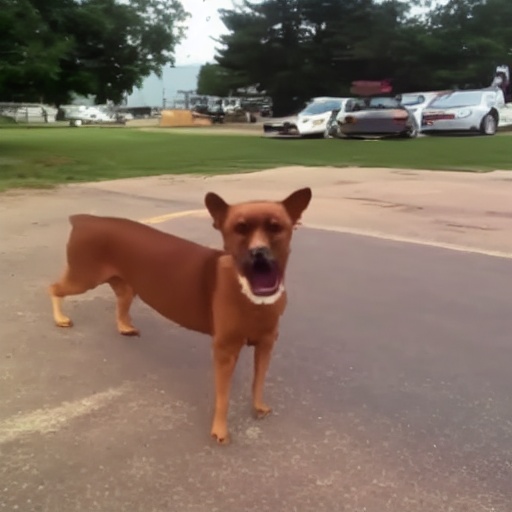}
        \label{c}
	\end{subfigure}
	\centering
	\begin{subfigure}
		\centering
		\includegraphics[width=0.18\linewidth]{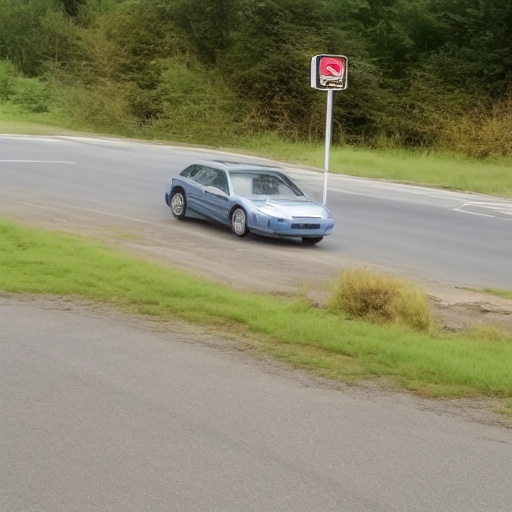}
	\end{subfigure}
        \centering
	\begin{subfigure}
		\centering
		\includegraphics[width=0.18\linewidth]{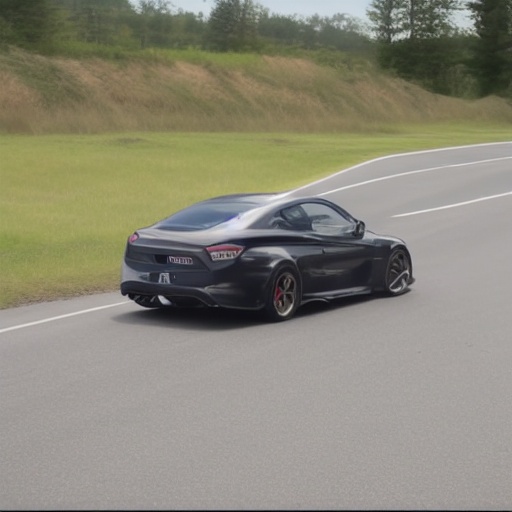}
	\end{subfigure}
	\caption{Generated images from interpolations between two audio clips (dog bark and motor vehicle).}
	\label{fig:interpolation}
\end{figure}

\textbf{Audios are Interpolable under MACS.} 
To evaluate the model’s ability to transition between sounds, we interpolate between the audio clips \(X\) (dog bark) and \(Y\) (motor vehicle, engine, revving):
\[
Z(\alpha) = \alpha X + (1 - \alpha)Y,\quad \alpha \in \{0,0.25,0.5,0.75,1\},
\]
where \(\alpha=0\) and \(1\) correspond to pure engine and bark.
At \(\alpha=0.5\), both dog and car appear. 
As shown in Fig.~\ref{fig:interpolation}, MACS successfully blends and disentangles semantic features from the mixed audio inputs.

\begin{figure}[ht]
    \centering
    \includegraphics[width=\linewidth]{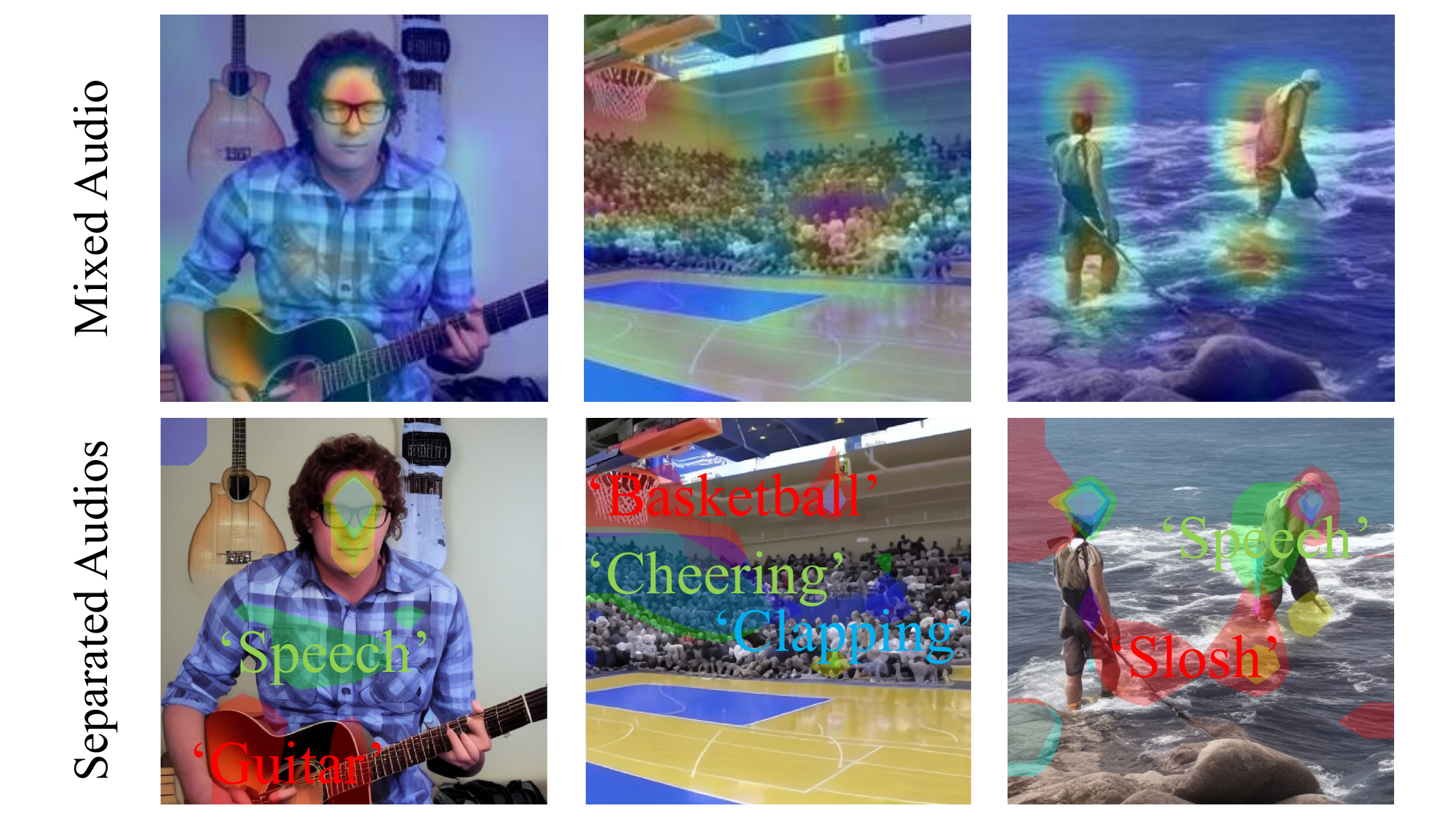}
    \caption{Attention maps of generated images w/ or w/o audio separation
    (Audio-text similarity score thresholded 0.5). Text colors indicate corresponding areas in the images.
    }
    \label{fig:attention}
\end{figure}

\textbf{Association of Separated Audios with Distinct Image Regions.}
\label{supp:attention}
We use Grad-CAM~\cite{selvaraju2017grad} to visualize how individual audio embeddings correspond to specific regions in the images generated by MACS. As shown in Fig.~\ref{fig:attention}, mixed audio embeddings result in diffuse attention maps, while disentangled embeddings produce more object-aligned regions, indicating stronger semantic alignment.


\subsection{Ablation Studies}
Due to space constraints, LLP-multi ablation results are reported in Appendix B. We ablated each of the three components, ranking loss (RL), contrastive loss (CL), and decoupled cross-attention (DC), individually, keeping the rest of MACS unchanged. In all cases, performance declined across all metrics, highlighting the importance of audio–text alignment for generating high-quality, semantically accurate images in multi-source audio-image generation.

\section{Conclusion}
We present MACS, the first two-stage architecture that explicitly separates mixed audio signals for audio-to-image generation. MACS preserves the contextual and semantic alignment between separated audio and text labels, and employs a decoupled cross-attention module to effectively integrate multiple audio inputs. Extensive experiments show that the ``separation before generation” strategy is effective, with MACS achieving state-of-the-art performance on both mixed- and single-source audio-to-image generation tasks.

\section*{Acknowledgements} 
This research is supported by the National Research Foundation, Singapore and Infocomm Media Development Authority under its Trust Tech Funding Initiative. Any opinions, findings and conclusions or recommendations expressed in this material are those of the author(s) and do not reflect the views of the National Research Foundation, Singapore and Infocomm Media Development Authority.

\bibliography{main}

\end{document}